# Extension of causal decomposition in the mutual complex dynamic process

**Author**
Yi Zhang[1,2,3*†], Qin Yang[1†], Lifu Zhang[1†,] Branko Celler[4], Steven Su[5], Peng Xu[2,3], Dezhong Yao[2,3]

*Correspondence: yi.zhang@uestc.edu.cn
† Yi Zhang, Qin Yang, and Lifu Zhang contributed equally to this work

1 School of Aeronautics and Astronautics, University of Electronic Science and Technology of China, Chengdu, Sichuan 611731, China
2 Key Laboratory for NeuroInformation of Ministry of Education, School of Life Science and Technology, University of Electronic Science and Technology of China, Chengdu, Sichuan 611731, China
3 Center for Information in BioMedicine, University of Electronic Science and Technology of China, Chengdu, Sichuan 611731, China
4 Biomedical Systems Laboratory, University of New South Wales, Sydney, New South Wales 2052, Australia
5 Centre for Health Technologies, Faculty of Engineering and Information Technology, University of Technology, Sydney, NSW 2220, Australia

**Abstract**

Causal decomposition depicts a cause-effect relationship that is not based on the concept of prediction, but based on the phase dependence of time series. It has been validated in both stochastic and deterministic systems and is now anticipated for its application in the complex dynamic process. Here, we present an extension of causal decomposition in the mutual complex dynamic process: cause and effect of time series are inherited in the decomposition of intrinsic components in a similar time scale. Furthermore, we illustrate comparative studies with predominate methods used in neuroscience, and show the applicability of the method particularly to physiological time series in brain-muscle interactions, implying the potential to the causality analysis in the complex physiological process.

**Introduction**

Since functional segregation and functional integration in neuroscience by Gall that distinct brain functions are localized in specialized cortical areas, the scientific criteria for assessing brain perception, cognition, and behavior have been dominated by Granger causality analysis via functional connectivity (FC) [1][2][3]. FC is used as a visual tool to show the temporal correlation among spatially distant brain regions [4][5][6].

Cause and effect in the most real-world are likely time-dependent, simultaneous, and reciprocal [7]. However, existing methods of the causality analysis in time series are predominantly based on the notion of statistical prediction, and may fail to describe the perspectives of simultaneity and reciprocity in causality philosophies. In this context, Granger causality is based on the assumption that cause and effect are separable, which is often satisfied in linear stochastic systems [8][9][10][11], but might not be applicable in complex dynamical processes (e.g., the brain-related physiological network) [12][13]. Moreover, the convergent cross-mapping (CCM) method introduces the state-space reconstruction to accommodate the inseparability of causal-effect [14]. In the causal modelling of brain network, dynamic causal modeling (DCM) [15][16][17] and transfer entropy (TE) [18][19][20][21] are also prevalent but essentially based on the Bayesian prediction.

Recently, Yang et al. propose a causal decomposition method based on the ensemble empirical mode decomposition (EEMD) for the causality analysis [7][22][23][24]. It follows the philosophical fundamental – the covariation principle of cause and effect, proposed by Galilei [25] and Hume [26]: variable A causes variable B if and only if, without variable A, variable B would not exist. Following this statement, the causal decomposition is not based on prediction, but based on the instantaneous phase dependency between cause and effect. It is assumed that variable A causes variable B

[25] if the instantaneous phase dependency between A and B is eliminated when the intrinsic mode functions (IMFs) in B is removed from B itself, but not vice versa. The study provides the adequate comparative analyses that validate the consistency of causal decomposition compared to Granger causality, CCM, and mutual information from mixed embedding, and exhibit its stability in causality inference [7].

The causal interaction is encoded in instantaneous phase dependency between two time series at a specific time scale [22]. To achieve this, the causal decomposition employs the generic EEMD to decompose a time series into a set of IMFs. As discussed by Huang and others, while each IMF represents a dynamic process in certain time scale, EEMD cannot guarantee the equal number of IMFs across two time series, and thus the mode alignment (cross-channel interdependence) revealing the similarity of time scales between two time series cannot work well [27-29]. Because the time series are often sampled in a distinct time scale in the complex dynamic process, EEMD-based causal decomposition perhaps is not directly applicable. Another constraint of the EEMD framework is that it is vulnerable to noise that may cause the appearance of the mode mixing (single-channel independence) [30]. Furthermore, time series are interestingly observed on the basis of a physiological network, often specifically including a complex process jointly coordinated by brain, cardiac-respiratory, neuromuscular activities within specific frequency bands [31]. To identify the causal interaction of such network, the application of EEMD-based casual decomposition is insufficient to find the intrinsic causal components.

Here, we present a noise-assisted multivariate EMD (NA-MEMD) causal decomposition analysis that is based on EEMD causal-decomposition, and more importantly, accommodates mode alignment and mode mixing, as well as offer an IMF selection method for the identification of the primary causal IMF pair between two time series [28][29][32][33]. We extend the theoretical work of causal decomposition to a complex dynamic process; that is, variable A causes variable B if the decomposed IMFs in both A and B are in certain similar time scale, and the instantaneous phase dependency between IMFs in A and B is eliminated when the IMF in B is removed from B itself, but not vice versa. To achieve this, we use the NA-MEMD to decompose two time series into IMFs, identify the primary causal IMF pair, and determine the causality between two time series. We validate the NA-MEMD causal-decomposition with both stochastic and deterministic systems and illustrate its application to physiological time series data of brain-muscle interactions in the movement of elbow extension and flexion.

## Results

**Experiment to brain-muscle causal interactions in the movement of elbow extension and flexion**

Figure 1 shows the subject experiment for acquiring the causality between brain and muscle in a typical elbow movement stimulation. The ethics committee in the University of Electronic Science and Technology of China (UESTC) approved the study, and the experiment is performed in the Key Laboratory of the Ministry of Neuro Information Education at UESTC. None of participants have a history of mental illness, neurological dysfunction, as well as are not currently taking psychoactive drugs. Twenty-two healthy untrained right-handed students from UESTC volunteer to join the experiment. The experimental environment is set in a darkened, electrical shielding room. A written informed consent is obtained from the participant who authorize the use of experiment data. Prior to the experiment, the experiment procedure is instructed to each individual (Fig. 1a). In a session, the experiment protocol (Fig. 1b) is repeated for the thirty times. E-prime is used to enhance the mental focus as well as guide the subject to follow the experiment protocol. The ANT Neuro eego$^{TM}$ mylab is utilized to simultaneously record electroencephalography (EEG) and electromyography (EMG) time series signals. A two-and-half-kilogram dumbbell is gripped during the experiment session (Fig. 1c). The EMG electrodes are placed on the midline of the biceps, and subjects are asked to consciously hold the elbow angle ($\theta$) at the horizontal and vertical positions at the respective flexion and extension movements (Fig. 1d).

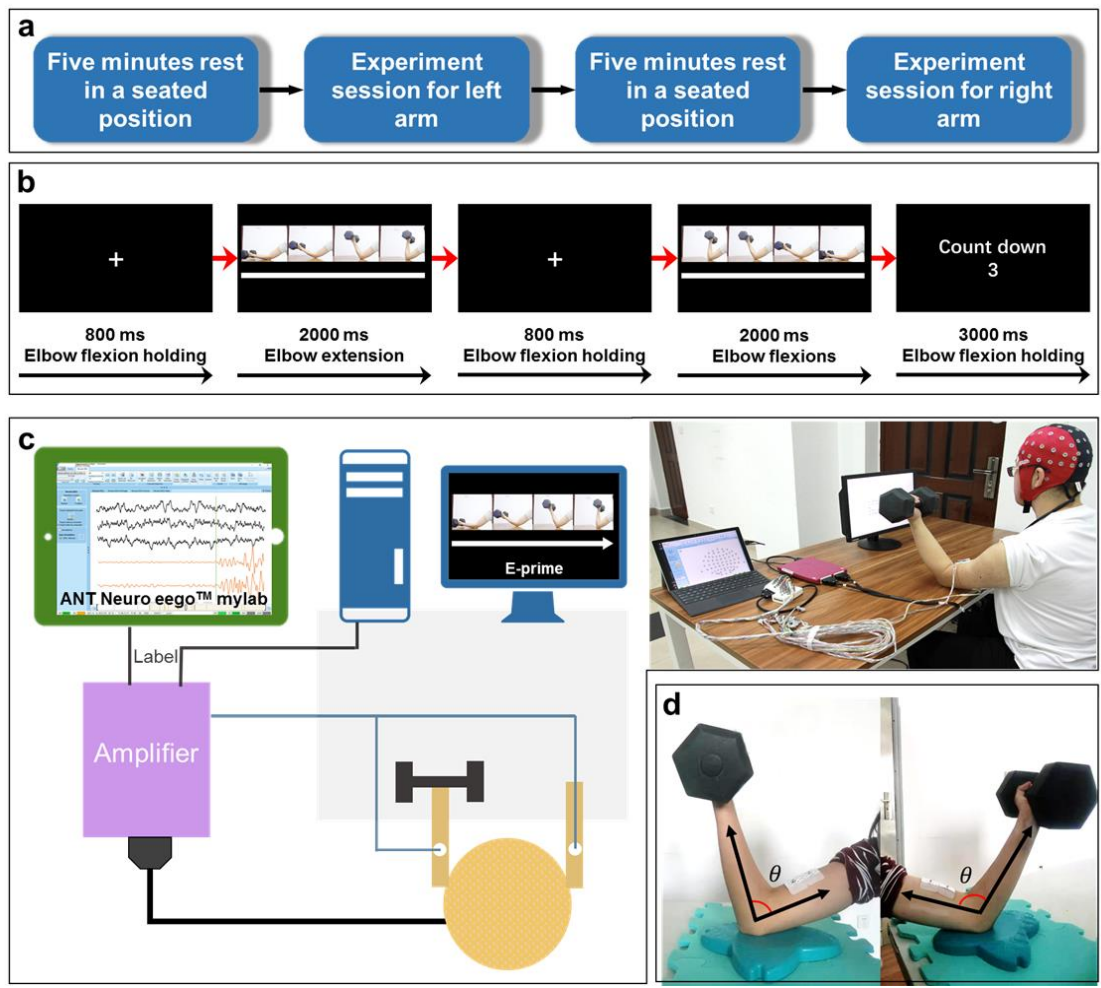

Figure 1: Experiment scenarios. **a** The experiment procedure for the illustration of EEG-EMG data acquisition representing a typical brain-muscle casual interaction. The time interval in approximate five minutes between sessions for each individual was allowed in order to avoid muscle fatigue. **b** The experiment protocol delivered by the E-prime system, **c** experiment settings and devices, **d** and illustrations of the weight standardized movement for the arm muscle activity are used in this case. The strategy of choosing a same weighted dumbbell in the experiment is to avoid the inter-subject variability. The sampling rate was set to 2000 Hz.

**Illustration of the NA-MEMD causal-decomposition method**

Figure 2 illustrates the NA-MEMD causal decomposition method used to identify the causal relationship of EEG and EMG time series.

Raw EEG and EMG time series are 50-Hz filtered to eliminate the power-line interference. The average reference is then used in EEG data. According to Welch's spectral density estimation, the band-pass filters are configured to 1-160 Hz and 20-350 Hz for EEG and EMG, respectively. EEG artifacts including EOG, and slight head or body movements, are rejected. After data pre-processing, EEG and EMG time series are decomposed into two set of IMFs using NA-MEMD (Fig. 2a). The selected IMFs (e.g., IMF5) are removed from EMG (Fig. 2b; subtract IMF 5 from the original EMG) and from EEG (Fig. 2c) and re-decomposed the time series. According to the instantaneous phase and instantaneous frequency of each IMF pairs, the removal of IMF 5 (i.e., Fig. 2d, e), was determined which minimizes the difference of the averaged instantaneous phases, and maximizes the average of instantaneous frequencies between the paired IMFs. The orthogonality and separability tests (Fig. 2f, g) are used to configure the NA-MEMD parameter (i.e., added noise level) for solving mode mixing and mode alignment. The casual strength (Fig. 2h) is computed showing the causality of original EEG and EMG signals, where a ratio approaching 1 indicated a strong causal influence from EEG signals to EMG signals.

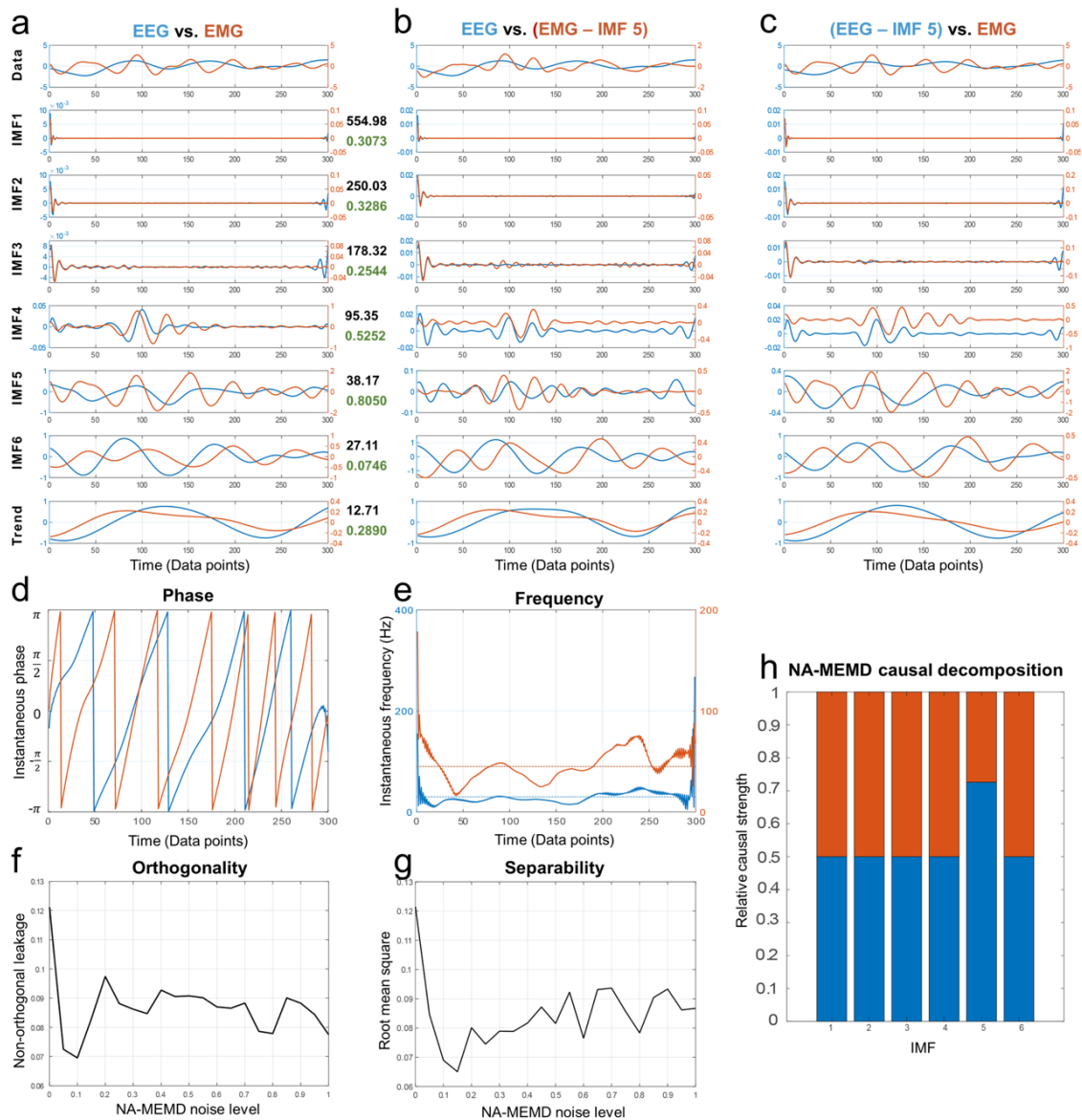

Figure 2: NA-MEMD causal decomposition analysis for EEG (blue line) and EMG (red line) time series. The EEG time series in the brain map location of C$_3$ (primary motor cortex), and the EMG time series on the biceps muscle are simultaneously collected. **a** Original time series are decomposed into six IMFs and one trend by using NA-MEMD. Each IMF in either EEG or EMG operated at distinct time scales. The averaged instantaneous frequency values (black) and the ratio of averaged instantaneous phase values (green) between paired IMFs are shown on the right side. **b** Time series with a removal of IMF 5 from EMG are re-decomposed by NA-MEMD. **c** Repeating the re-decomposition process in the EEG time series. **d** The instantaneous phase and **e** the instantaneous frequency plotting. **f** Nonorthogonality and **g** separability tests. In this case, the noise level (λ) at 0.1 standard deviations of the time series was compromised as both orthogonality and separability tests maintained acceptable outcomes. **h**

Generalization of NA-MEMD causal decomposition to each IMF.

**Validation of NA-MEMD causal-decomposition analysis**

Figure 3 validates the causal decomposition with existing methods by assessing the effects of down-sampling and temporal shift for the deterministic [6][7], stochastic [7][34], and EEG-EMG processes (Fig. 3a). The effect of the down-sampling on NA-MEMD and EEMD based causal decompositions, and Granger causality are comparatively assessed (Fig. 3b). Those original time series are down-sampled by a down-sample factor (from 1 to 10). Since the down-sampling damages the causal dynamics, the causality inference became difficult in the prediction based Granger causality analysis. Compared to outcomes in the casual decomposition, either EEMD causal decomposition or NA-MEMD causal decomposition can offer a sound effect in deterministic, stochastic, and EEG-EMG systems when down-sampling factor is less than 2, following a stabilized outcome with no causality effect. Fig. 3c shows the results of assessing the NA-MEMD causal decomposition to temporal shift with a lag-lead transformation up to 20 data points. The NA-MEMD causal decomposition exhibited a stable pattern of causal strength in both lag and lead shifts. The time shift does not generate a spurious causal relationship. The EEMD causal decomposition also showed stable causal strengths in the deterministic system, but resulted in opposite causal strengths in stochastic and EEG-EMG systems. Moreover, Granger causality lost its predictability with inconsistent results in temporal shift for deterministic, stochastic, and EEG-EMG systems. Fig. 3d illustrates the performance of NA-MEMD causal decomposition in uncorrelated white noise data with varying lengths (L∈[10, 1000]). NA-MEMD causal decomposition presented a consistent pattern of causal strengths at an approximate value of 0.5 (the shadow areas indicated the standard deviation at 20 repetitions in terms of the data length). In the case of Granger causality, there was no significance, indicating no causality. However, EEMD causal decomposition detected the spurious causality from the beginning up to 300 data points.

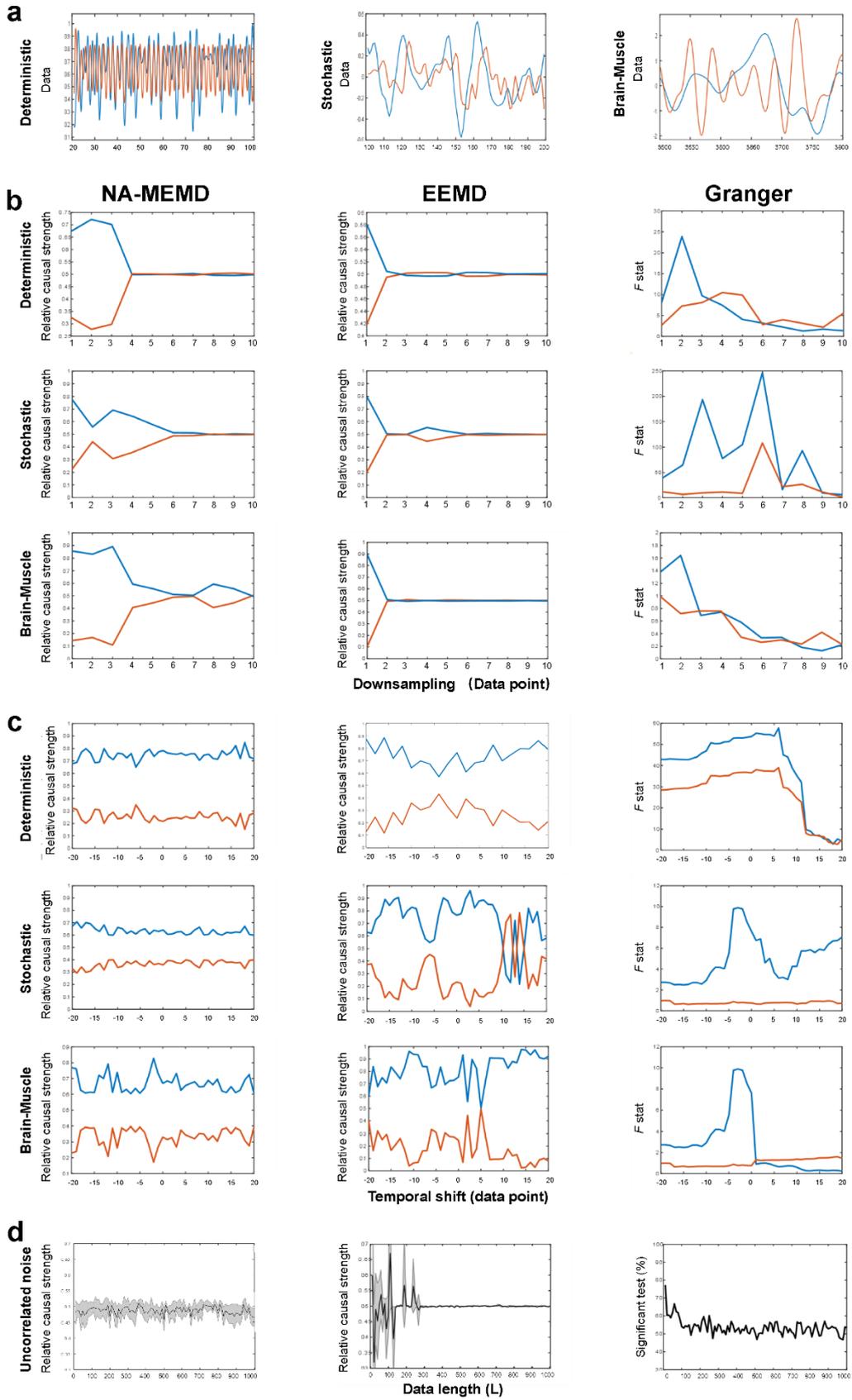

Figure 3: Deterministic, stochastic and EEG-EMG model evaluation.
**a** Comparative studies of NA-MEMD casual decomposition to deterministic system (left), stochastic system (middle) and EEG-EMG system (right). **b** Comparison of down-sampling effects to NA-MEMD casual decomposition, EEMD casual decomposition and Granger causality. Evaluations for **c** temporal shift, and **d** data length effects of uncorrelated white noise time-series. Data length: 80 data points (deterministic system); 100 data points (stochastic system); 300 data points (EEG-EMG system).

**Application of NA-MEMD causal decomposition in the brain-muscle dynamic process**

Figure 4 shows the application of NA-MEMD causal decomposition for exploring the causal dynamics in a typical motor behavior (Fig. 1). From the perspective of the accumulation of cause, the cause observing from $C_3$ ($E_1$) is larger than that from biceps muscle at time windows 50, 100, 200. Considering the neuron activation of biceps muscle in unexecuted upper limbs, the accumulation of cause ($E_2$) in the executed limb is larger than that in unexecuted one.

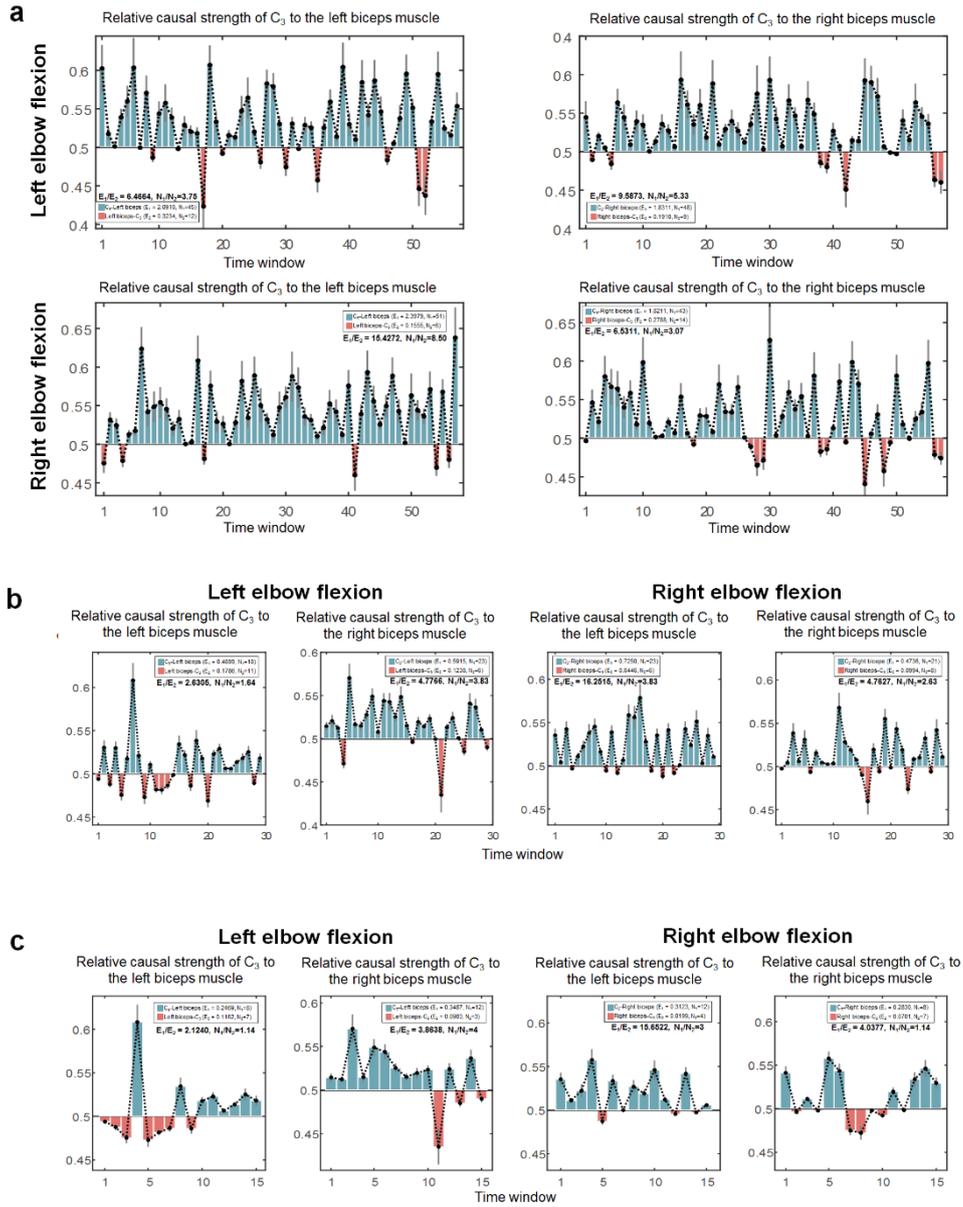

Figure 4: Application of the NA-MEMD causal decomposition method to identify the causal relationship between brain and biceps muscle to elbow movement. Experiment data with twenty-two subjects are used to explore the dynamics of causality in terms of time window, $C_3$-biceps muscle (movement executed), and $C_3$-biceps muscle (movement unexecuted). The two-second elbow flexion EEG-EMG data of all subjects with thirty-times repetitions are segmented to **a** 50, **b** 100, **c** 200-point time windows. Statistical results (mean ± STD with 95% confidence interval) are assessed based on $C_3$-left/right biceps muscles with and without the motion execution. The relative causal strength based on time window is acquired. The accumulated causal effects are described by the sum of the averaged relative causal strengths of time windows, where

$E_1$ represents the comprehensive causality from $C_3$ to biceps muscle, $E_2$ meaning that from biceps muscle to $C_3$. The predominant causal-effect role between $C_3$ and biceps muscle is depicted by counting up the number of $C_3$-biceps muscle ($C_3$: cause, biceps muscle: effect, $N_1$) and biceps muscle-$C_3$ ($N_2$).

**Discussion**

Our work contributes to extend the EEMD causal decomposition method to the mutual dynamic complex process. The physiologic system or brain activity has attracted considerable attention for exploring the coordinated causal relationship in perception, cognition, and behavior. However, when considering the brain-related physiological system, the time series bio-signals across different time scales are the key issue that may fail to the practice of EEMD causal decomposition.

The fundamentals of causal decomposition is based on the inference of causality proposed by Galilei and Hume. NA-MEMD exhibits the covariation principle with a capacity of revealing time-dependency, simultaneity and reciprocity of cause and effect. Those properties in a complex dynamic process should be inherited through its decomposition of intrinsic causal components, namely as IMF. Both EEMD and NA-MEMD casual decomposition methods are similarly based on causality philosophies. In addition, the prediction based Granger's causality contains the prior knowledge of limited time period which is insufficient to infer causality since the history might be biased already.

Previous research confirmed that the general problem related to mode-alignment and mode-mixing would affect the pattern in decomposed IMFs [28][35]. It is therefore unlikely to use EEMD for two and multiple time series [36]. For instance, EEMD cannot guarantee the equal number of IMFs in different time series. This also can result in that a nominated IMF pair is not often in a similar time scale [37][38]. Therefore, the EEMD based phase coherence analysis is not acceptable for the causality inference of a complex physiological process.

Based on previous studies on the physiological system (e.g., EMG) in multiple time series channels, NA-MEMD provides an equal number of IMFs across different channels with less overlap of information among the decomposed IMFs in each EMG channel. In other words, NA-MEMD outperforms than EEMD for mode-alignment and mode-mixing [39][40][28][29], and is effective for simultaneously analyzing IMFs based time scales. It was consistent with the results in Fig. 3, where the decomposition of time series with NA-MEMD presented more stable characteristics in the deterministic, stochastic and EEG-EMG systems.

Both the EEMD and NA-MEMD methods are based on the Galileo's definition, which are based on the instantaneous phase dependency. The spectral extension of Granger's causality is based on time dependency, and prone to be influenced by non-stationary artefacts. The results in Fig 3b. and Fig 3c. demonstrate that phase dependency based methods with more stable causality pattern (v.s., time dependency based Granger causality).

This work is fundamentally limited by several factors: (1) similar with EEMD based causal decomposition and granger causality, NA-MEMD causal decomposition is also a reprehensive of statistical causality, which is not a true causality; (2) the parameter determination for the noise-assisted data analytics is vital, which is also important for a unbiased result in the NA-MEMD decomposition; (3) the definition of phase-dependent causality for two time series is merely suitable for two time series, and not applicable for the causal analysis of multivariable time series.

The typical brain-related limb movement experiment can show the general utility of NA-MEMD causal decomposition method in coupled nonlinear chaotic oscillators. Certainly, the results in deterministic and stochastic systems can also validate the consistency of NA-MEMD and EEMD causal decompositions with true causality. The exploration of NA-MEMD method expands the scope of phase-dependent causality analysis, implying the potential to the complex dynamic systems such as the physiological network.

**Methods**

**Noise-assisted multivariate empirical mode decomposition**
Assume that there are two time series $u_1(t), u_2(t)$ and $m$ uncorrelated white gaussian noise time series $g_1(t), g_2(t), \ldots, g_m(t)$. Define a matrix X as the resulting multiple time series given by

$$X = \begin{bmatrix} u_1(t) \\ u_2(t) \\ g_1(t) \\ g_2(t) \\ \vdots \\ g_m(t) \end{bmatrix} \quad (1)$$

Let $P = \{p_1, p_2, \ldots, p_{m+1}\}$ be the first $(m+1)$ prime point set. Define the $i^{th}$ sample of a one-dimensional Halton sequence $d_i^P$, then we have the following equivalent direction vector equations:

$$d_i^{p_\theta} = \frac{\alpha_0}{p_\theta} + \frac{\alpha_1}{p_\theta^2} + \frac{\alpha_2}{p_\theta^3} + \cdots + \frac{\alpha_\rho}{p_\theta^{\rho+1}} (\theta \in [1, m+1]) \quad (2)$$

$$i = \alpha_0 + \alpha_1 \times p_\theta + \alpha_2 \times p_\theta^2 + \cdots + \alpha_\rho \times p_\theta^\rho \quad (3)$$

where $\rho$ is given by the representation of $i$ mirrored at the decimal point, and $\alpha_0, \alpha_1, \ldots, \alpha_\rho$ is the individual digits of $i$ in $p_\theta$-binary representation.

The Hammersley sequence is then defined and the $i^{th}$ direction vector set is then obtained as $h_i = (\frac{i}{N}, d_i^{p_1}, d_i^{p_2}, \ldots, d_i^{p_{m+1}})$ and $i \in \{1, 2, \cdots, N\}, N \in \mathbb{N}^+$.

In terms of the low-discrepancy Hammersley sequence in equation (2)-(3), the Hammersley point matrix $H_s$ in prime number representation can be rewritten as

$$H_s = \begin{bmatrix} h_1 \\ h_2 \\ \vdots \\ h_N \end{bmatrix} \qquad (4)$$

Define unit direction matrix $r_i$ on $(m+2)$-sphere which can be computed by $h_i$. Then the projection of the resulting multiple time series on $(m+2)$-sphere can be depicted by

$$Y = r_i X^T \qquad (5)$$

Following EMD and equation (5), the minima and maxima of $Y$ are interpolated to obtain the respective lower and upper envelope curves. Then Intrinsic Mode Functions (IMFs) $C_{\xi, \hat{m}}$ can be given by

$$y_{\hat{m}}(t) = \sum_{\zeta=1}^{N} c_{\zeta, \hat{m}}(t) + r(t) \quad (\hat{m} = 1, 2) \qquad (6)$$

where $c_{\zeta, 1}$, $c_{\zeta, 2}$ are called IMF pair.

**Definition (Phase-dependent Causality for Two Time Series).** The two time series $u_1, u_2$ defined in equation (1) is said to be a cause-effect relationship if the decomposed IMF pair in equation (6) satisfies the following conditions: (1) the absolute difference of the instantaneous phase of IMF pair is negligible small. (2) the mean frequency of IMF pair derived via Hilbert-Huang transform (HHT) is maximal. (3) the IMF pair with respect to relative causal strength $(C)$ is said to be a cause-effect relationship.

**Remark** In the above definition, the two cause-effect time series can have the similar or non-similar time scale. Condition (1) indicates a boundary constraint given by

$$|mean(\phi_1(t)) - \gamma \, mean(\phi_2(t))| < \delta \qquad (7)$$

where $\phi_1(t)$, $\phi_2(t)$ are the instantaneous phase via Hilbert-Huang transform (HHT). $mean(\cdot)$ is time averaging, $\delta$ is a constant, and $\gamma \approx 1$.

Assume that IMF can be described as $c_{\zeta, \hat{m}} = a_\zeta(t) \cos(\phi_{\hat{m}}(t))$ where $a(t)$ is instantaneous amplitude and $\phi(t)$ is instantaneous phase, Define the IMF pair

$\{\mathcal{F}_1, \mathcal{F}_2\}$, IMF based Hilbert spectrum set $\hat{H} = \{\widetilde{H_1}, \widetilde{H_2}, \ldots, \widetilde{H_\zeta}\}$ and IMF based Hilbert spectrum $\widetilde{H_k} = \int_0^T (H_{1,k} + H_{2,k})$   $k = 1, 2, \ldots \zeta$, where $H_{i,k} = \int_0^T \sum_{\widehat{m}=1}^2 H(a,t)dt, i = 1, 2$. The IMF pair $\mathcal{F}_{1,2}$ can be determined if and only if the corresponding IMF pair satisfies the following equation,

$$\{\mathcal{F}_1, \mathcal{F}_2\} <=> max(\hat{H}) \tag{8}$$

Then the causality of $\{\mathcal{F}_1, \mathcal{F}_2\}$ represent the causality of $\{u_1, u_2\}$. The phase coherence between IMF pair $\{\mathcal{F}_1, \mathcal{F}_2\}$ is given by

$$\eta(\mathcal{F}_1, \mathcal{F}_2) = \frac{1}{T}\left|\int_0^T exp(j(\phi_1(t) - \phi_2(t))) \cdot dt\right| \tag{9}$$

where $\phi_1, \phi_2$ is the instantaneous phase computed by Hilbert transform. Then absolute causal strengths can be computed by

$$\begin{aligned}\sigma(u_2 \to u_1) &= |\eta(u_1, u_2) - \eta((u_1 - \mathcal{F}_1), u_2)| \\ \sigma(u_1 \to u_2) &= |\eta(u_1, u_2) - \eta(u_1, (u_2 - \mathcal{F}_2))|\end{aligned} \tag{10}$$

The relative causal strength $C$ is defined as the relative ratio of the two absolute causal strengths, where $C \in [0,1]$.

$$\begin{aligned}C(u_1 \to u_2) &= \frac{\sigma(u_1 \to u_2)}{\sigma(u_1 \to u_2) + \sigma(u_2 \to u_1)} \\ C(u_2 \to u_1) &= \frac{\sigma(u_2 \to u_1)}{\sigma(u_1 \to u_2) + \sigma(u_2 \to u_1)}\end{aligned} \tag{11}$$

The ratio of 0.5 implies that there was either no causality or reciprocal causality. A ratio of 0 or 1 indicates a strong differential causal influence from $u_1$ or $u_2$.

**Deterministic, stochastic data** The deterministic model was used in accordance with Sugihara et al. [6] based on a coupled two-species nonlinear logistic difference system, expressed as follows (initial value x(1) = 0.2, and y(1) = 0.4).

$$X(t + 1) = X(t)[3.8 - 3.8X(t) - 0.02Y(t)] \tag{12}$$

$$Y(t + 1) = Y(t)[3.5 - 3.5Y(t) - 0.1X(t)] \tag{13}$$

For stochastic system, the part of example shown in Ding et al. [41] for Granger causality were used, which is expressed as follows (using a random number as the initial value).

$$X(t + 1) = 0.95\sqrt{2}X(t) - 0.9025X(t - 1) + w_1(t) \tag{14}$$

$$Y(t + 1) = 0.5X(t - 1) + w_2(t) \tag{15}$$